\documentclass[aps,pre,twocolumn,preprintnumbers,amssymb,showpacs,superscriptaddress,floatfix]{revtex4-1}
\usepackage{amstext,amssymb}
\usepackage{amsmath}
\usepackage{amsfonts}
\usepackage{xcolor}
\usepackage{graphicx}
\usepackage{dcolumn}
\usepackage{bm}
\usepackage{color}
\usepackage{bbold}

\newcommand{\ini}{p_{in}}
\newcommand{\an}{d}
\newcommand{\cre}{d^{\dagger}}
\newcommand{\rate}{T}

\begin{document}

\title{Kinetics and thermodynamics of a driven open quantum system}

\author{Juzar Thingna}
\affiliation{Complex Systems and Statistical Mechanics, Physics and Materials Science Research Unit, University of Luxembourg, L-1511 Luxembourg, Luxembourg}
\author{Felipe Barra}
\affiliation{Departamento de F\'isica, Facultad de Ciencias F\'isicas y Matem\'aticas, Universidad de Chile, 837.0415 Santiago, Chile}
\author{Massimiliano Esposito}
\email[]{massimiliano.esposito@uni.lu}
\affiliation{Complex Systems and Statistical Mechanics, Physics and Materials Science Research Unit, University of Luxembourg, L-1511 Luxembourg, Luxembourg}

\date{\today}

\begin{abstract}
Redfield theory provides a closed kinetic description of a quantum system in weak contact with a very dense reservoir. Landau-Zener theory does the same for a time-dependent driven system in contact with a sparse reservoir. Using a simple model, we analyze the validity of these two theories by comparing their predictions with exact numerical results. We show that despite their \emph{a priori} different range of validity, these two descriptions can give rise to an identical quantum master equation. Both theories can be used for a nonequilibrium thermodynamic description which we show is consistent with exact thermodynamic identities evaluated in the full system-reservoir space. We emphasize the importance of properly accounting for the system-reservoir interaction energy and of operating in regimes where the reservoir can be considered as close to ideal.   
\end{abstract}

\maketitle

\section{Introduction}\label{sec:1}

Solving the full dynamics of an open quantum system in contact with a large reservoir is often very hard. The aim of kinetic theory is to derive a closed description for the open system dynamics by tracing out the reservoir degrees of freedom \cite{Spohn80, Breuer2002, Weiss2008}. Several kinetic schemes have been proposed for both driven and autonomous systems. The Redfield master equation is commonly used and relies on weak system-reservoir coupling \cite{Redfield1957}. In case of autonomous systems, it has been tested against exactly solvable models. A key requirement for it to hold is to have a sufficiently strong coupling between the system and the reservoir to induce an effective mixing between the system and reservoir eigenstates \cite{Esposito2003, Esposito2003b}, a requirement closely related to the celebrated eigenstate thermalization hypothesis \cite{Alessio16}.

The validity of kinetic schemes for time-dependently driven open quantum systems is rarely explored. Recently, this issue was addressed for driven open systems interacting with sparse reservoirs using a Landau-Zener treatment of the system-reservoir level crossing dynamics and which leads to a discrete time Markov chain evolution for the system occupation \cite{BarraPRE16}. It was shown to match very well with the exact quantum dynamics of a linearly driven spinless quantum dot interacting with a sparse reservoir. Building on this work, we now explore the different dynamical regimes of this model. In particular, when moving from sparser to denser reservoir densities, the validity of the Landau-Zener and Redfield kinetic schemes is critically assessed and compared with exact quantum simulations. 

In the dense reservoir regime, we find that the time-dependent driving acting on the system improves the validity of the Redfield scheme compared to the autonomous case. The reason is that the driving effectively increases the number of reservoir levels with which the system interacts, although it does so sequentially instead of simultaneously. In the sparse regime, the Landau-Zener based description seem to hold even beyond its regime of validity. Remarkably, when a continuous time limit of the Landau-Zener scheme is justified, the resulting master equation takes the same form as the Redfield quantum master equation although the latter is not supposed to hold in these regimes.

We then explore the thermodynamic description in the different kinetic regimes and show that they are perfectly compatible with the exact thermodynamics identities previously derived in the full system-reservoir space in Ref.~\cite{Esposito2010}. For the Redfield theory, our results show the importance of including the system-reservoir interaction energy in the system when going beyond the Born-Markov-Secular approximation, a result consistent with other recent findings in Refs.~\cite{Barra2015, Strasberg2017}. In the Landau-Zener regime, we show that while coherent effects are important in the full system dynamics, the discrete time description from one system-reservoir crossing to the next is insensitive to them. Interesting insight is also obtained on the validity of the ``ideal'' reservoir assumption needed for the kinetic and thermodynamics description to hold, in particular in relation to the positivity of the entropy production rate.

The paper is organized as follows: In Sec.~\ref{sec:2} we present the model and describe the numerical approach used to evaluate the exact quantum dynamics. In Sec~\ref{sec:3} we first presents the two kinetic schemes, the Redfield one for dense reservoirs and the Landau-Zener one for sparse reservoirs. We then assess their validity by comparison with the exact quantum dynamics. In Sec~\ref{sec:4} we first present the thermodynamics based on the two kinetic schemes and then compare it with exact thermodynamic identities numerically evaluated. The case of periodic driving is discussed in Sec.~\ref{sec:5}. Conclusions and perspectives are drawn in Sec.~\ref{sec:6}.

\section{Model and exact dynamics}\label{sec:2}

\begin{figure}[tb!]
\includegraphics[width=0.8\columnwidth]{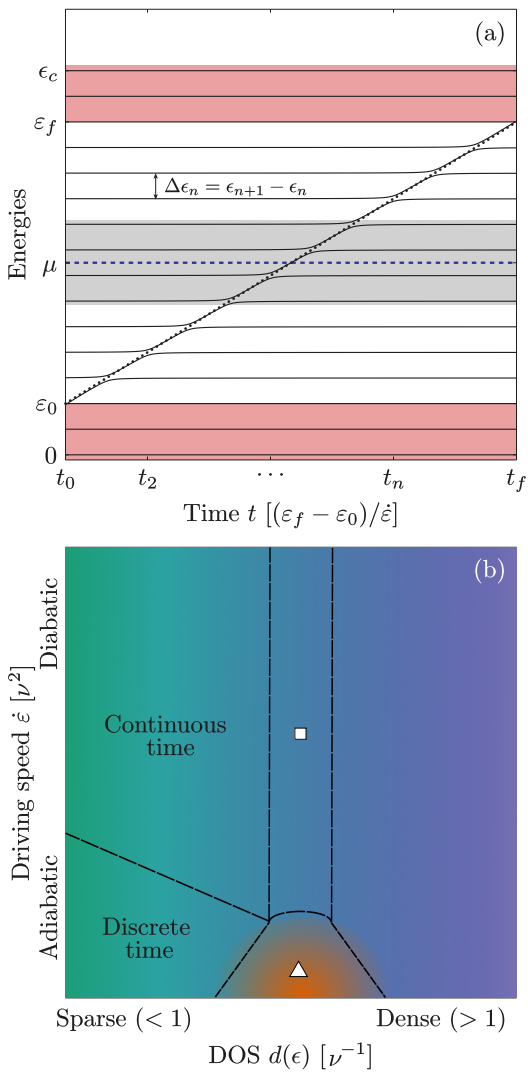}
\caption{\label{fig1} (Color Online) Panel (a) illustrates the spectra of the model as a function of time. The bare energy of the driven quantum dot (black dotted line) is linearly ramped in time while the bare reservoir level do not change. The interacting levels are essentially identical to the bare ones except when the dot level crosses a reservoir level, resulting in an avoided crossing. The gray region symbolically represents the region of order $k_B T$ around the chemical potential $\mu$ (blue dashed line) where the reservoir occupation is significantly different from one or zero. The red region represents a region that the dot level never penetrates in order to avoid edge effects from the density of state. Panel (b) schematizes the regions of different relaxation for a given system-reservoir coupling $\nu$. The time-dependent Redfield (Landau-Zener) description holds in the violet (green) region, while no known kinetic description holds in the central orange region. The triangle and the square indicate where the parameters of Fig.~\ref{fig4}(a) and Fig.~\ref{fig4}(b) lie.}
\end{figure}

We consider the following many body-Hamiltonian
\begin{eqnarray}
\label{eq:1}
H(t) &=& \varepsilon_t \cre\an + \sum_{n=1}^{L}\epsilon_n c_n^{\dagger}c_n + \sum_{n=1}^{L} \big (\nu_n \cre c_n + \mathrm{H.c.} \big),\\
&\equiv & H_S(t) + H_R + V, \nonumber
\end{eqnarray}
which can be thought of as a driven single-level (spinless) quantum dot interacting with a reservoir made of $L$ non-driven quantum dots. The onsite energy of the system is time-dependent like in the case of single-electron pumps \cite{Jauho94, Switkes99}. The system and reservoir are assumed uncorrelated initially, $\rho(0)=\varrho(0) \otimes \rho_R^{eq}$. The initial condition for the system is diagonal in the system energy basis whereas the reservoir is at equilibrium in a grand-canonical distribution $\rho_R^{eq} = \exp [-\beta (H_R-\mu N_R)]/Z_R$ with $Z_R$ being the grand partition function of the reservoir and $N_R = \sum_{n=1}^{L} c_n^{\dagger}c_n$ the particle number.

Since we consider diagonal initial states $\varrho(0)$ and the many-body Hamiltonian [Eq.~(\ref{eq:1})] is non-interacting, the dynamics is fully captured by the single-body Hamiltonian 
\begin{eqnarray}
\label{eq:2}
H^1(t) &=& \varepsilon_t |\varphi_0\rangle\langle\varphi_0 | + \sum_{n=1}^{L}\epsilon_n |\varphi_n\rangle\langle\varphi_n |  \nonumber \\
&&+ \sum_{n=1}^{L} \big( \nu_n |\varphi_n\rangle\langle\varphi_0 | + \mathrm{H.c.} \big),
\end{eqnarray}
where $|\varphi_i\rangle$ is the single body basis of dimension $L+1$ diagonalizing the bare single particle system and reservoir Hamiltonian (i.e. $H^1$ is diagonal when all the $\nu_n=0$). The corresponding single-body time-evolution operator reads $U^1(t,t')=\exp\left[-i\int_{t'}^t d\tau H^1(\tau)\right]$, where time-ordering is not needed since the Hamiltonian commutes with itself at different times $[H(t),H(t')]=0$. The occupation probability of the dot at time $t$ will for instance be given by 
\begin{eqnarray}
\label{eq:3}
p(t) &=& \mathrm{Tr}\left[\cre\an\rho(t)\right]=\langle\varphi_0 | U^{1}(t,0)\rho^1(0) U^1(0,t) | \varphi_0\rangle .
\end{eqnarray}
The initial single-body density matrix $\rho^{1}(0)$ is a diagonal matrix whose first element is the initial occupation of the dot and the $L$ remaining ones are the Fermi-Dirac distribution $f(\epsilon_n) = \{\exp[\beta(\epsilon_n - \mu)] +1\}^{-1}$ describing the occupation of the reservoir level $n$. Above $\mathrm{Tr}^{1}$ indicates the trace over the reservoir degrees in the single-body space.

The dynamics of this system can be simulated exactly and will be used as a reference result later on to be compared with. For the exact simulations we will always assume that the reservoir levels are equally spaced, $\epsilon_n=n\epsilon_c/L$, and we choose all system-reservoir coupling terms constant $\nu_n=\nu~\forall ~n$. The time dependent driving of the system energy is chosen to be linear $\varepsilon_t = \varepsilon_0 + \dot{\varepsilon} t$. An illustration of the total system energetics is provided in Fig.~\ref{fig1}(a). The discretization scheme used to perform the numerical evolution makes use of the implicit mid-point rule that calculates the exponential of the Hamiltonian by first order P\`{a}de approximation \cite{Crank}:
\begin{eqnarray}
\label{eq:4}
U^1(t+\Delta t,t) &\approx & \frac{\mathbb{1}-i\frac{\Delta t}{2}H^1(t+\frac{\Delta t}{2})}{\mathbb{1}+i\frac{\Delta t}{2}H^1(t+\frac{\Delta t}{2})}.
\end{eqnarray}
Consequently, the propagator from time $0$ to time $t$ is given by $U^1(t,0)=\Pi_{i=1}^{N}U^1(t_i+\Delta t,t_i)$ where $t_1 = 0$, $t_{N+1}=t$. Throughout this work we will set $\hbar$ (Planck constant) $=k_B$ (Boltzmann constant) $= e$ (the electron charge) $= 1$. 

\section{Effective system dynamics}\label{sec:3}

We now turn to two different regimes for which a closed effective master equation dynamics can be obtained for the system. In both cases the validity of these kinetic descriptions will be assessed by comparison with the numerically exact full quantum dynamics.

\subsection*{Redfield master equation (RQME)}\label{subsec:3.1}

We first consider the regime where the Redfield master equation describes the system's dynamics \cite{Breuer2002, Redfield1957}. This description is valid for systems weakly interacting with large and dense reservoirs ($L \to \infty$). But in practice, when comparing to numerically exact simulation where the reservoir has a discrete spectra, one finds that the system-reservoir coupling cannot be too weak. It must be sufficiently strong such that the coupling energy range contains many reservoir levels, but is nevertheless small as compared to the energy scale of variation of the reservoir density of state \cite{Esposito2003, Esposito2003b}. For short, we will say that the interaction needs to be strong enough to mix the reservoir levels \cite{note1}. While the traditional Redfield theory is for autonomous Hamiltonians, it can be easily extended to time-dependent system Hamiltonians \cite{Zhou2015, Bulnes16}. In this case we are not aware of attempts to compare Redfield theory to a numerically exact full system dynamics.  

The Redfield quantum master equation \cite{Breuer2002, Redfield1957, Zhou2015, Gaspard1999} for our model, Eq.~(\ref{eq:1}), reads 
\begin{eqnarray}
\label{eq:5}
d_t\varrho_{nm} = \sum_{i,j}\mathcal{L}^{ij}_{nm}\varrho_{ij}, 
\end{eqnarray}
where the non-zero elements of $\mathcal{L}^{ij}_{nm}$ are given by
\begin{eqnarray}
\mathcal{L}^{11}_{11} &=& -\left(\rate_{21}^{12}+\rate_{21}^{12*}\right), \quad \mathcal{L}^{12}_{12} = -\left(\rate_{21}^{12}+\rate_{12}^{21*}\right)+i\varepsilon_t, \nonumber \\
\mathcal{L}^{22}_{11} &=& \rate_{12}^{21}+\rate_{12}^{21*}, \quad \mathcal{L}^{21}_{21} = -\left(\rate_{12}^{21}+\rate_{21}^{12*}\right)-i\varepsilon_t,\nonumber \\
\mathcal{L}^{11}_{11} &=&-\mathcal{L}^{11}_{22} \,, \quad \mathcal{L}^{22}_{11}=-\mathcal{L}^{22}_{22}
\end{eqnarray}
and $\varrho(t) = \mathrm{Tr}_R[\rho(t)]$ is the reduced density matrix of the driven system. Throughout this work the explicit derivative will be denoted by $d_x$ with $x$ being the variable of differentiation. The simplicity of our model implies that coherences and populations dynamically decouple without the need to invoke the rotating wave approximation. This Redfield equation is non-Markovian \cite{Gaspard1999} due the explicit time-dependence of its coefficients
\begin{eqnarray}
\label{eq:6}
\rate_{12}^{kl}(t)&=&\int_{0}^{t}dt'e^{i\int_{0}^{t'}\varepsilon_{t'-\tau}d\tau}C^{kl}(t'),\nonumber \\
\rate_{21}^{kl}(t)&=&\int_{0}^{t}dt'e^{-i\int_{0}^{t'}\varepsilon_{t'-\tau}d\tau}C^{kl}(t'),
\end{eqnarray}
expressed in terms of the reservoir correlation functions 
\begin{eqnarray}
\label{eq:7}
C^{12}(t')&=&\int_{-\infty}^{\infty}\frac{d\epsilon}{2\pi}\Gamma(\epsilon)f(\epsilon)e^{i\epsilon t'}, \\
\label{eq:8}
C^{21}(t')&=&\int_{-\infty}^{\infty}\frac{d\epsilon}{2\pi}\Gamma(\epsilon)[1-f(\epsilon)]e^{-i\epsilon t'},
\end{eqnarray}
where $f(\epsilon) = \{\exp[\beta(\epsilon - \mu)] +1\}^{-1}$ denotes the Fermi-Dirac distribution. The spectral density $\Gamma(\epsilon) = 2 \pi\sum_{n=-\infty}^{\infty}|\nu_n|^2\delta(\epsilon-\epsilon_n)$ encodes information about both, the density of state $d(\epsilon) = \sum_{n=-\infty}^{\infty}\delta(\epsilon-\epsilon_n)$, and the system-reservoir couplings $\nu_n$. The dense reservoir condition enters at this point. The spectral density and the density of state are replaced by their smoothed version in the Redfield description. The smoothed density of state $\bar{d}(\epsilon)$ gives the number of states within an energy shell of width $\delta_\epsilon$ centered at $\epsilon$, divided by $\delta_\epsilon$. The width should be large compared to the level-spacing and small compared to energy variations at a classical scale. Similarly, the smoothed spectral density is given by $\bar{\Gamma}(\epsilon)=2 \pi \nu^2(\epsilon)\bar{d}(\epsilon)$ where the coupling elements $\nu_n$ are supposed to behave as a smooth function of the energy $\nu(\epsilon)$. The evolution of the system occupation, $\varrho_{22}(t) \equiv p(t)$, specifically reads
\begin{eqnarray}
\label{eq:10}
d_tp(t) &=& \rate^{+}(t)[1-p(t)]-\rate^{-}(t)p(t),
\end{eqnarray}
where the non-Markovian transition rates are given by $\rate^{+}(t) = 2\mathrm{Re}[\rate^{12}_{21}]$ and $\rate^{-}(t) = 2\mathrm{Re}[\rate^{21}_{12}]$. For driven systems, the Markovian approximation describing the dynamics over timescales longer than the typical decay time of the reservoir correlation functions $C^{ij}(t)$, $\tau_r$, relies on the following assumptions. First, the driving is very slow compared to $\tau_r$ so that $\varepsilon_{t'-\tau}$ can be replaced by $\varepsilon_{t'}$ under the integral in Eq.~(\ref{eq:6}). Second, since the correlation functions in Eq.~(\ref{eq:6}) decay to zero for times larger than $\tau_r$, their time argument can be extended to infinity. One easily verifies that in this case 
\begin{eqnarray}
& \rate^{+}(\infty)= \Gamma(\varepsilon_t) f(\varepsilon_t) \nonumber \\
& \rate^{-}(\infty) = \Gamma(\varepsilon_t) [1-f(\varepsilon_t)]. \label{FermiRates}
\end{eqnarray}
We note that these rates satisfy local detailed balance, namely $\rate^{+}(\infty)/\rate^{-}(\infty)=\exp{\lbrack -\beta (\varepsilon_t-\mu)\rbrack}$. 
   
\begin{figure}
\includegraphics[width=\columnwidth]{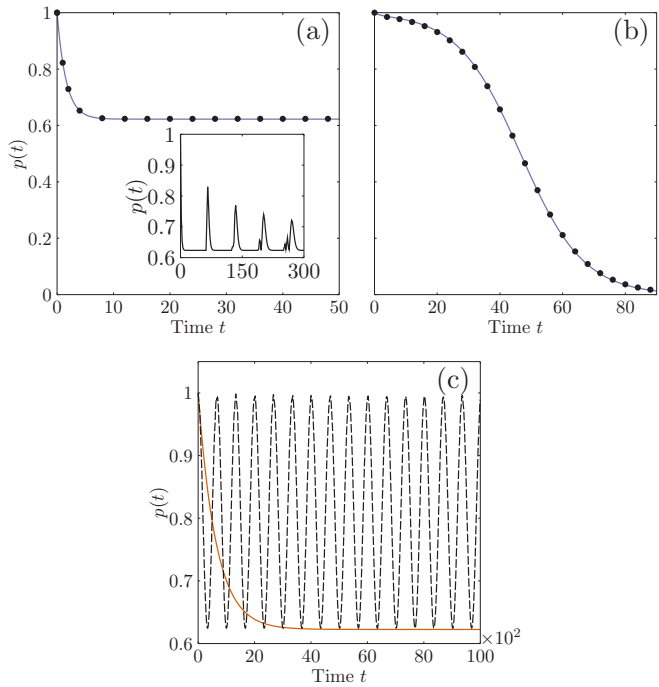}
\caption{\label{fig2} (Color Online) Comparison between the system occupation predicted by the numerically exact full quantum dynamics [Eq.~(\ref{eq:3})] (black closed circles/dashed lines) and the non-Markovian Redfield equation [Eq.~(\ref{eq:10})] (violet solid lines). Since $L=1000$ and $\epsilon_c = 100$, then $\bar{d}=10$. In panel (a) and (b) the mixing condition $d \nu \gg 1$ is satisfied with $\nu=0.1$, whereas in panel (c) this condition is violated with with $\nu=0.005$. There is no driving in panels (a) and (c); $\varepsilon_t=5$ and $\mu=10$, while in panel (b) $\varepsilon_t=5+t$ with $\mu=50$. The inset in panel (a) depicts the longer time evolution that displays quantum recurrences. The inverse temperature $\beta=0.1$.}
\end{figure}

In Fig.~\ref{fig2}, we compare the non-Markovian RQME, Eq.~(\ref{eq:10}), to the numerically exact dynamics. To evaluate Eq.~(\ref{eq:7}) and Eq.~(\ref{eq:8}) we used the smoothed density of state and spectral density for our model given by $\bar{d}=L/\epsilon_c$ and $\bar{\Gamma}= 2 \pi \bar{d}\nu^2$ inside the interval $[0,\epsilon_c]$ and by zero outside. 
We note that the equivalence between the microcanonical and canonical version of the Redfield equation is guaranteed by the constant reservoir density of state implying an infinite reservoir microcanonical heat capacity \cite{Esposito2003, Esposito2007}.
The shortest time scale in the dynamics is the reservoir correlation time, $\tau_r \sim 1/\epsilon_c$, and the longest one (in absence of driving) is the Heisenberg recurrence time $\tau_h \sim d$ beyond which the kinetic description will fail. The relaxation time scale $\tau_{rel} \sim 1/\bar{\Gamma}$ must therefore be shorter then the $\tau_h$. Indeed, this means that $\bar{d} \nu \gg 1$, which is the condition for the interaction to mix the reservoir levels.
The numerically exact occupation [Eq.~(\ref{eq:3})] (black closed circles) is very well reproduced by the Redfield equation, in absence [Fig.~\ref{fig2}(a)] as well as in presence of driving [Fig.~\ref{fig2}(b)] as long as $\bar{d} \nu \gg 1$. When this condition is violated [Fig.~\ref{fig2}(c)], the exact dynamics displays large oscillations not captured at all by the Redfield dynamics. The inset of Fig.~\ref{fig2}(a) shows the quantum recurrences in the exact dynamics which arise after $\tau_{h}$. In presence of driving these are pushed much further away and thus not shown. 

\subsection*{Landau-Zener master equation (LZQME)}\label{subsec:3.2}

We now consider the regime where a Landau-Zener Markov chain recently proposed in Ref.~\cite{BarraPRE16} describes the dynamics. The validity of this approach relies on two key conditions: First the time needed between two avoided crossings, $\tau_c(\epsilon_n) = |\Delta\epsilon_n/\dot{\varepsilon}|$ where $\Delta \epsilon_n=\epsilon_{n+1}-\epsilon_n$ is the energy spacing between successive reservoir levels, must be long compared to the time needed for a crossing, $\tau_{ac}(\epsilon_n) = 2 \nu_n / |\dot{\varepsilon}|$ where $2\nu_n$ is the energy gap induced by the interaction between the levels. This condition ensures that the dynamics can be treated as a sequence of successive avoided level crossings between the driven system level and the reservoir levels. 
Second the time needed between two avoided crossings, $\tau_c(\epsilon_n)$, must also be long compared to the Landau-Zener time, $\tau_{lz}(\epsilon_n)\equiv|\dot{\varepsilon}|^{-1/2}\mathrm{max}(1,2\nu_n/|\dot{\varepsilon}|^{1/2})$, after which the Landau-Zener transition probabilities holds \cite{NoriPR2010}. This condition is meant to ensure that we can use Landau-Zener theory to evaluate transition probabilities. 
Using the first condition to simplify the second, we get that
\begin{eqnarray}\label{LZcond}
1) \; \; |\Delta\epsilon_n| > 2|\nu_n| \ \ \;, \ \ 2) \; \; |\Delta\epsilon_n| > \sqrt{|\dot{\varepsilon}|}.
\end{eqnarray}

The Landau-Zener probability of diabatic transition when the system level undergoes an avoiding crossing with a reservoir level is given by \cite{Landau1932, Zener1932, Majorana1932, Stueckelberg1932, DemkovJETP68}
\begin{eqnarray}
R_n = \exp [-2\pi |\nu_n|^2/\dot{\varepsilon}], 
\end{eqnarray}
where $\dot{\varepsilon}$ the driving rate at the crossing. In between two crossings the system occupation does not change. This means that the system occupation, $p_{n+1}$, just after crossing with level $n+1$ at time $t_{n+1}$ and having crossed level $n$ at energy $\epsilon_n$ and time $t_n$, can be written as a Markov chain 
\begin{eqnarray}
\label{eq:11}
p_{n+1} = R_n p_n + (1-R_n) f_n,
\end{eqnarray}
where $f_n=f(\epsilon_n)$ is the Fermi-Dirac distribution of the $n$th reservoir level. 
Equation~(\ref{eq:11}) is the discrete time Landau-Zener Markov chain recently proposed in Ref.~\cite{BarraPRE16}.

We now turn to the continuous-time version of this equation.
We consider a small time interval $dt$ during which $n$ levels are crossed
\begin{eqnarray}
\label{eq:11b}
\frac{p_n-p_0}{dt}= \frac{1}{dt} \sum_{l=0}^{n-1} (1-R_l) (f_l-p_l). 
\end{eqnarray}
Assuming that we can make $dt$ sufficiently small to be treated infinitesimally and neglect the variation of $(1-R_l) (f_l-p_l)$ under the sum, we can approximate this equation by 
\begin{eqnarray}
\label{eq:11c}
d_t p(t) = \dot\varepsilon \bar{d}(\varepsilon_t) [1-R(\varepsilon_t)] [f(\varepsilon_t)-p(t)],
\end{eqnarray}
where $\bar{d}(\varepsilon_t)$ is the continuous version of $1/\vert \Delta \epsilon_n \vert$ and $\dot\varepsilon \bar{d}(\varepsilon_t)$ an estimation of $n$.
\begin{figure*}[tb!]
\includegraphics[width=\textwidth]{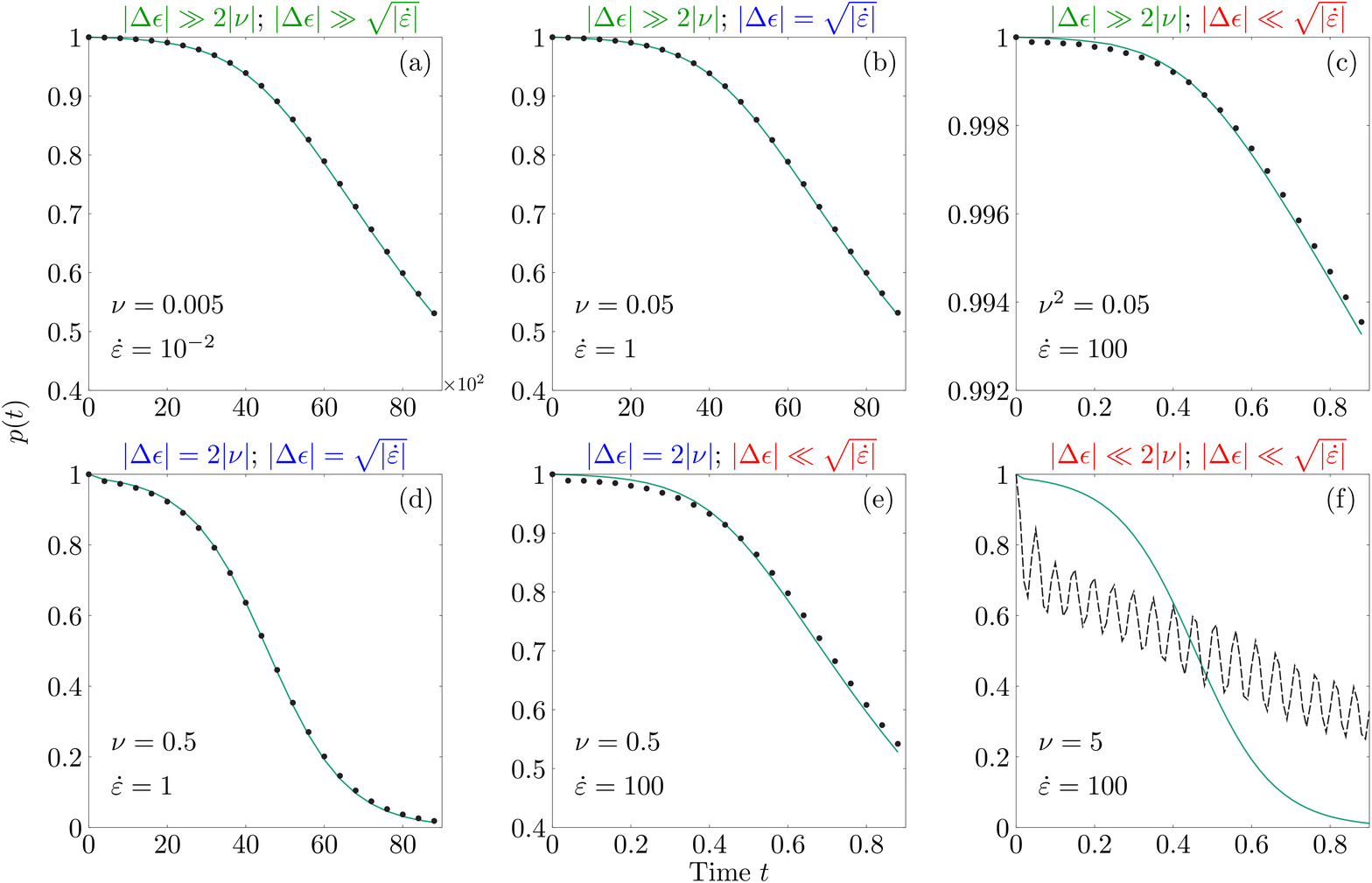}
\caption{\label{fig3} (Color Online) Comparison between the system occupation predicted by the numerically exact full quantum dynamics [Eq.~(\ref{eq:3})] (black closed circles/dashed lines) and the continuous-time Landau-Zener equation [Eq.~(\ref{eq:15})] (green solid line). The discrete Landau-Zener Markov chain perfectly matches with latter and is thus not shown. The driving is $\varepsilon_t=5+\dot{\varepsilon}t$ and the time axis is always chosen such that the system energy is linearly ramped from $\varepsilon_0=5$ to $\varepsilon_f=95$. Since $L = 100$ and $\epsilon_c = 100$, then $\vert \Delta \epsilon_n \vert=\vert \Delta \epsilon \vert=1$. Also $\beta = 0.1$ and $\mu = 50$.}
\end{figure*}
In the diabatic regime where $\dot{\varepsilon} \gg 2 \pi |\nu(\varepsilon_t)|^2$, we get  
\begin{eqnarray}
\label{eq:12}
d_t p(t) = 2 \pi \bar{d}(\varepsilon_t) |\nu(\varepsilon_t)|^2 [p(t)-f(\varepsilon_t)],
\end{eqnarray}
which is the Landau-Zener master equation 
\begin{eqnarray}
\label{eq:15}
d_t p(t) = \rate^{+}(\infty) [1-p(t)] - \rate^{-}(\infty)p(t),
\end{eqnarray}
i.e. a Pauli equation \cite{Pauli} with Fermi-golden rule rates $\rate^{\pm}(\infty)$ given by Eq.~(\ref{FermiRates}) and which depend on time via the energy of the system. Remarkably, Eq.~(\ref{eq:15}) is identical the Markovian version of the Eq.~(\ref{eq:10}), despite the fact that the former requires $\nu(\epsilon) \bar{d}(\epsilon)$ to be much smaller then one while the latter requires it to be much larger then one. Also, the physics governing the two regimes is quite contrasting. In the Landau-Zener regime the system level is moved across the reservoir levels and interacts sequentially with them, while in the Redfield regime the same level is continuously interacting with a large number of reservoir levels.

Fig.~\ref{fig3}(a) displays various dynamics of the dot occupation for the model described in Sec.~\ref{sec:2}, where $\bar{d}(\varepsilon_t)=L/\epsilon_c$ and $\nu_n=\nu~\forall ~n$. The dynamics is always in the diabatic regime and the discrete and continuous LZQME always perfectly match each other. The LZQME accurately captures the dynamics not only if conditions 1) and 2) are satisfied but also when one of them is violated [Fig.~\ref{fig3}(b) to (e)]. In this latter case, small discrepancies can be observed on short time scales, most evident in panels (c) and (e). When both conditions, 1) and 2), are violated [Fig.~\ref{fig3}(f)], the LZQME prediction fails to reproduce the exact quantum dynamics. 

We have shown that the exact quantum dynamics displays two regimes that can be accurately described by kinetic equations. The LZQME describes the low reservoir density of state regime while the RQME describes the high reservoir density of state regime [respectively the left and the right part of Fig.~{\ref{fig1}(b)]. Remarkably, if the driving is fast enough, the (continuous) LZQME becomes identical to the Markovian RQME. In case of low reservoir density of state, we expect that the discrete LZQME is valid everywhere and the LZQME is valid only when the driving is fast (diabatic regime). When the reservoir density of state is very low we expect that the LZQME (continuous time) would be valid if and only if the driving speed compensates for the sparseness of the reservoir. As the density increases, still being in the sparse regime, the driving speed required for the validity of LZQME would decrease [as depicted by the dashed boundary line for continuous time LZQME in Fig.~{\ref{fig1}(b)}]. 

In the intermediate (orange) part of Fig.~{\ref{fig1}(b), the exact quantum dynamics of the dot can display more complex behaviors as can be seen in Fig.~\ref{fig4}. For slow driving speed (panel a) the exact quantum dynamics is irregular on intermediate times and is clearly not captured by any quantum master equation. However, the long time behavior coincides with the RQME (or the discrete LZQME) because the multiple crossings of reservoir levels generated by the driving eventually produce on the dot an effect analogous to the mixing of the reservoir levels needed for the RQME to work. For higher driving speed (panel b), this effect is more pronounced and the driving compensates for the lack of mixing at any time [hence depicted by the mixing of violet (RQME) and green (LZQME) colors in Fig.~{\ref{fig1}}(b)]. We can thus conclude that multiple reservoir level crossings caused by driving, and mixing of reservoir levels caused by the system-reservoir interaction, have effectively a similar role and enable the system to relax.  
\begin{figure}[t!]
\includegraphics[width=\columnwidth]{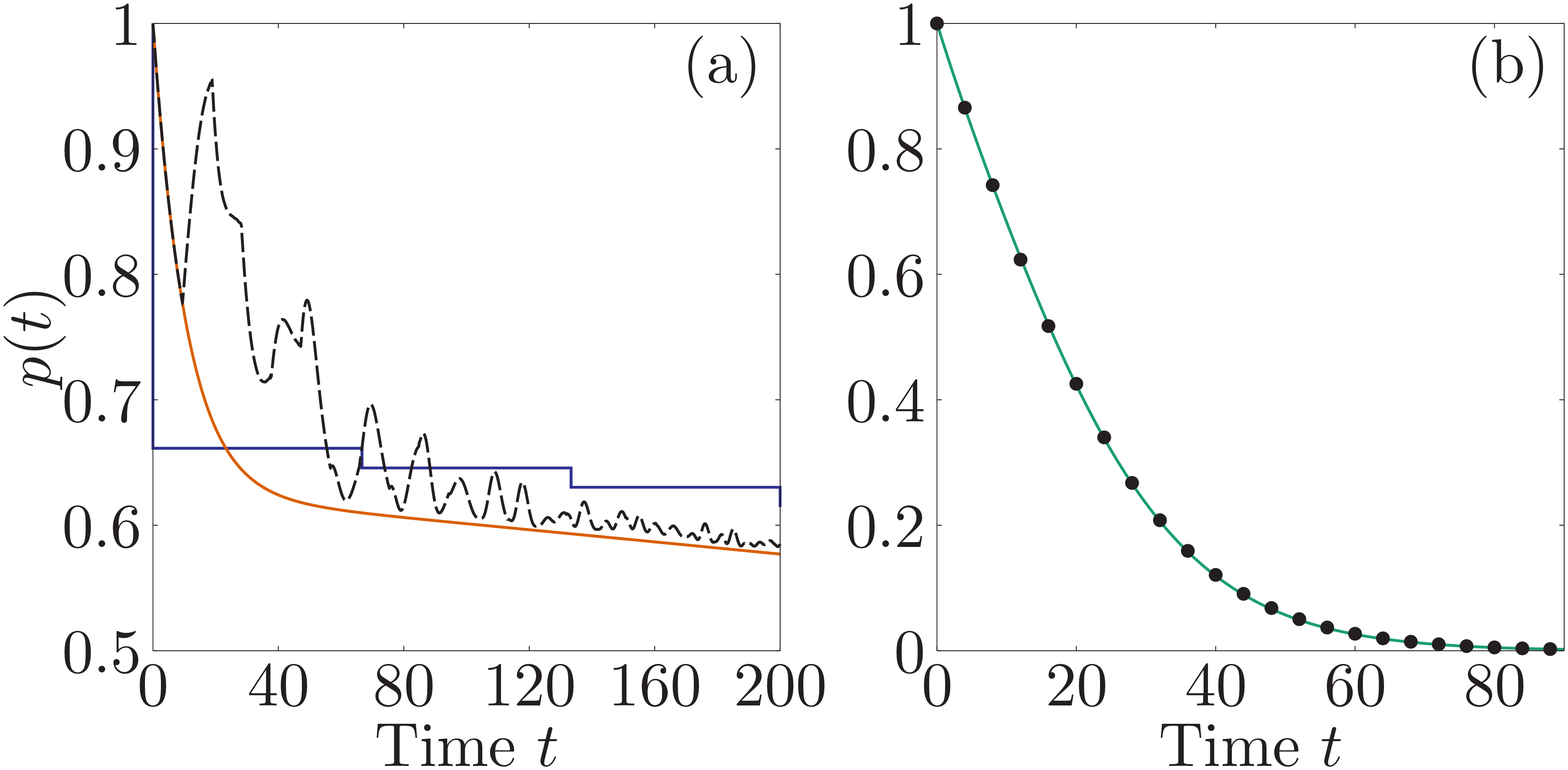}
\caption{\label{fig4} (Color Online) System occupation corresponding to intermediate reservoir density of state [central region of Fig.~\ref{fig1}(b)]. The driving is the same as in Fig.\ref{fig3}. The exact dynamics is denoted by a dashed line in panel (a) where the driving is slow, $\dot{\varepsilon}=0.01$, and black closed circles in panel (b) where the driving is moderate, $\dot{\varepsilon}=1$. The Markovian RQME $\equiv$ LZQME is denoted by an orange and green solid line, while the discrete LZQME is denoted by a blue solid line. $L=150$, $\epsilon_c = 100$, $\nu = 0.1$, $\beta = 0.1$, and $\mu = 10$.}
\end{figure}


\section{Thermodynamics}\label{sec:4}

We now turn to the thermodynamic description. We investigate the equivalence between thermodynamic quantities such as heat, work, and entropy production defined in the total system \cite{Esposito2010} and their corresponding expression in terms of the kinetic description.

\subsection*{Total system description}\label{subsec:4.1}

Following Refs.~\cite{Esposito2010, Strasberg2017}, the heat and work rate of an open system Eq.~(\ref{eq:1}) are defined by
\begin{eqnarray}
\label{eq:18}
&&\dot{Q} = I_E - \mu I_N,\\
\label{eq:19}
&&\dot{W} = \dot{W}_{mech} + \mu I_N, 
\end{eqnarray}
where the energy current, the particle current and the mechanical work due to the time-dependent driving are given by 
\begin{eqnarray}
\label{eq:20}
&&I_E = -\mathrm{Tr}\left[d_t \rho(t) H_R\right], \\
\label{eq:21}
&&I_N = -\mathrm{Tr}\left[d_t \rho(t) N_R\right],\\
\label{eq:22}
&&\dot{W}_{mech} = \mathrm{Tr}\left[d_t H_S(t)\rho(t)\right],
\end{eqnarray}
and $\mu I_N$ is the chemical work. We note that dotted quantities are by definition rates.  
Considering that the system entropy is the von Neumann entropy of the system $S(t)= -\mathrm{Tr}_S[\varrho(t)\mathrm{ln}\varrho(t)]$,
one finds both, the first law of thermodynamics 
\begin{eqnarray}
\label{eq:23}
U(t)-U(0)=Q+W \equiv \int_0^t d\tau (\dot{Q}+ \dot{W}) 
\end{eqnarray}
where the system energy is given by
\begin{eqnarray}
\label{eq:24}
U(t) &=& \mathrm{Tr} \left[\rho(t)\left\{H_S(t)+V\right\}\right],
\end{eqnarray}
as well as the second law of thermodynamics
\begin{eqnarray}
\label{eq:25}
\Delta_i S = D\left[\rho(t)||\varrho(t)\otimes\rho_R^{eq}\right] = S(t)-S(0) - \beta Q \geq 0, \nonumber \\
\end{eqnarray}
where $\rho_R^{eq}$ is the initial grand-canonical distribution of the reservoir defined below Eq.~(\ref{eq:1}). 
The entropy production is non-negative because it can be expressed as a quantum relative entropy $D[\rho || \rho'] = \mathrm{Tr}[\rho\mathrm{ln}\rho-\rho'\mathrm{ln}\rho']\geq 0$. 

\subsection*{Time-dependent Redfield regime}\label{subsec:4.2}

Next, we provide the thermodynamic expressions for the currents, Eqs.~(\ref{eq:20}) and (\ref{eq:21}), that can be explicitly derived within the same approximation as the non-Markovian RQME (following Refs.~\cite{Zhou2015, Bulnes16, Esposito2009, Thingna2012, Thingna2014}) and expressed solely in terms of quantities appearing in the RQME
\begin{eqnarray}
\label{eq:26}
&&I_E = \bar{\rate}^{+}(t)[1-p(t)]-\bar{\rate}^{-}(t)p(t), \\
\label{eq:27}
&&I_{N} = \rate^{+}(t)[1-p(t)]-\rate^{-}(t)p(t), \\
\label{eq:28}
&&\dot{W}_{mech} = \dot{\varepsilon}_t p(t),
\end{eqnarray}
with $\rate^+(t)$ and $\rate^{-}(t)$ defined below Eq.~(\ref{eq:10}) while
\begin{eqnarray}
\bar{\rate}^{+}(t) &=& 2\mathrm{Im}\left[\int_{0}^{t}dt'\exp\left(-i\int_{0}^{t'}\varepsilon_{t'-\tau}d\tau\right)d_{t'}C^{12}(t')\right],\nonumber\\
\label{eq:29}
\bar{\rate}^{-}(t) &=& 2\mathrm{Im}\left[\int_{0}^{t}dt'\exp\left(i\int_{0}^{t'}\varepsilon_{t'-\tau}d\tau\right)d_{t'}C^{21}(t')\right].
\end{eqnarray}
The system entropy is in turn given by $S(t)=-p(t)\mathrm{ln}p(t)-[1-p(t)]\mathrm{ln}[1-p(t)]$ due to the lack of coherences. We can now use these expressions to calculate every thermodynamic quantity within Redfield theory using the relations given in Sec.~\ref{subsec:4.1}. 

Let us note that a common formulation of the first law for quantum master equations is 
\begin{eqnarray}
\label{eq:30}
d_t\mathcal{U} &=& \mathrm{Tr}\left[d_t\rho(t)H_S(t)\right]+\mathrm{Tr}\left[\rho(t)d_tH_S(t)\right].\nonumber\\
&\equiv& \dot{\mathcal{Q}}+\dot{\mathcal{W}} \nonumber \\
&=& (\varepsilon_t-\mu) d_t p(t)+\left[\dot{\varepsilon}_t p(t)+\mu d_t p(t)\right].
\end{eqnarray}
This formulation neglects the contribution of the coupling $V$ in the energy [see Eq.~(\ref{eq:24})]. The discrepancy between the two approaches only plays a role in the non-Markovian regime. Indeed, in the Markovian limit, via integration by parts one can show $\bar{\rate}^{\pm}(\infty)=\varepsilon_t\rate^{\pm}(\infty)$, implying that in this limit $\dot{\mathcal{Q}} \equiv \dot{Q}$ and $\dot{\mathcal{W}} \equiv \dot{W}$. This agreement in the Markovian limit is not general but specific to our model, or to any model where coherences and populations in the system eigenbasis are decoupled (e.g. in the rotating wave or secular approximation). The general statement is that the thermodynamic formulation based on Eq.~(\ref{eq:30}) and the Redfield equation is consistent only in the Markov-Secular approximation.

\subsection*{Landau-Zener regime}\label{subsec:4.3}

In the Landau-Zener regime, the first and second law are given by \cite{BarraPRE16},
\begin{eqnarray}
&&U(t_{n+1})-U(t_{n}) = W(t_{n+1},t_{n})+Q(t_{n+1},t_{n}), \\
&&\Delta_i S(t_{n+1},t_{n}) = S(t_{n+1})-S(t_{n}) - \beta Q(t_{n+1},t_{n}) \geq 0, \nonumber
\end{eqnarray}
where
\begin{eqnarray}
&&U(t_n)=\varepsilon_t p_n \ \ \;, \ \ N(t_n)= p_n, \\
&&S(t_n)=-p_n \mathrm{ln}p_n-[1-p_n]\mathrm{ln}[1-p_n],  \\
&&W_{mech}(t_{n+1},t_{n}) = (\epsilon_{n+1}-\epsilon_n)p_{n+1}, \\
&&Q(t_{n+1},t_{n}) = (\epsilon_{n}-\mu)[p_{n+1}-p_n].
\end{eqnarray}
In the continuous time limit, the thermodynamic quantities can be obtained from 
\begin{eqnarray}
\label{eq:31}
&&I_E = \varepsilon_t\left[\rate^{+}(\infty)[1-p(t)]-\rate^{-}(\infty)p(t)\right], \\
\label{eq:32}
&&I_N = \rate^{+}(\infty)[1-p(t)]-\rate^{-}(\infty)p(t), \\
\label{eq:33}
&&\dot{W}_{mech} = \dot{\varepsilon}_t p(t).
\end{eqnarray}
Not surprisingly, these expressions coincide with the Markovian limit of the RQME Eq.~(\ref{eq:30}), since the continuous time LZQME coincides with the Markovian RQME.

\subsection*{Numerical comparison}\label{subsec:4.4}

\begin{figure}[tb!]
\includegraphics[width=\columnwidth]{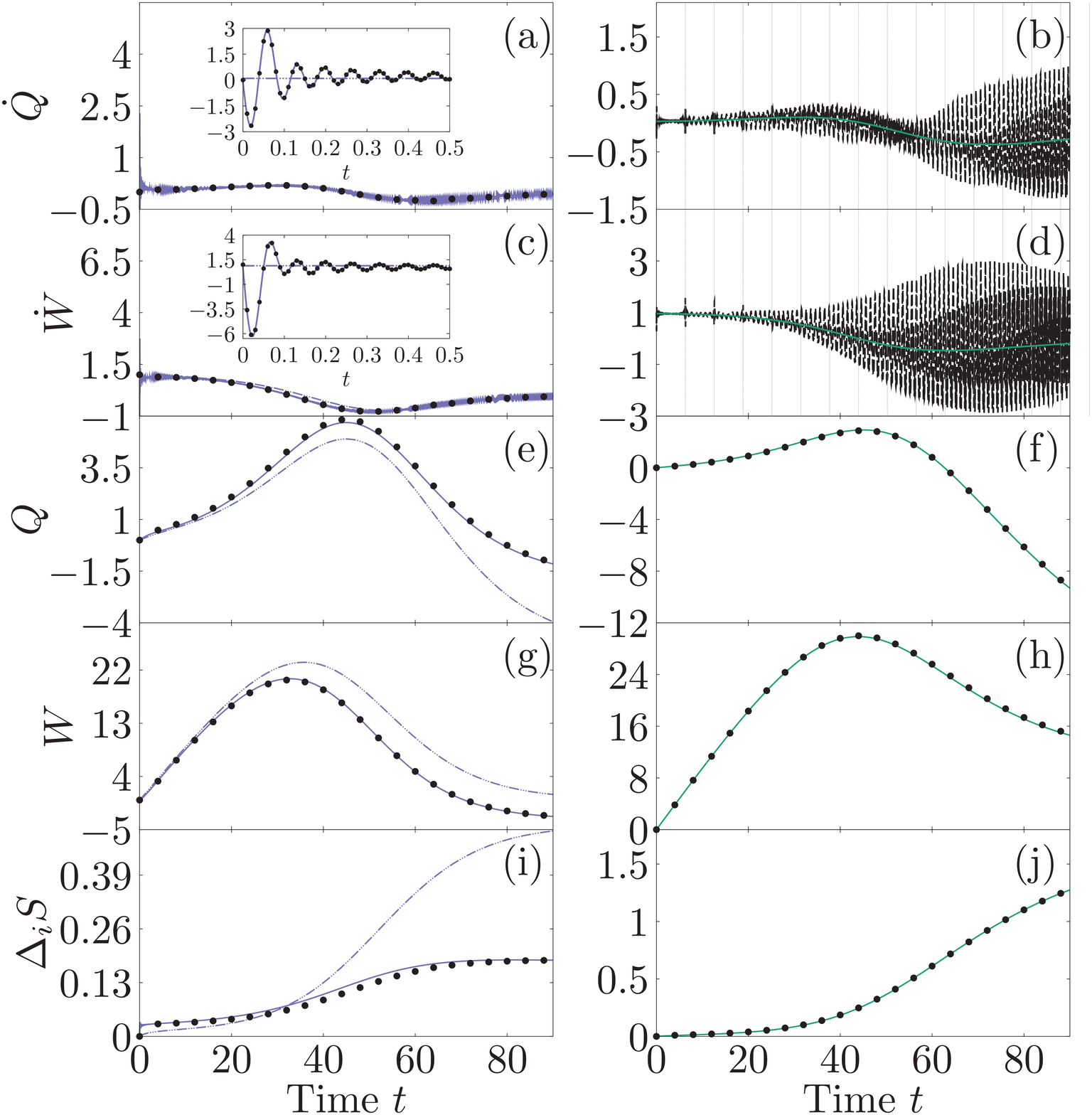}
\caption{\label{fig5} (Color Online) Heat rate $\dot{Q}$ [panels (a) and (b)], work rate $\dot{W}$ [panels (c) and (d)], heat $Q$ [panels (e) and (f)], work $W$ [panels (g) and (h)], and entropy production $\Delta_i S$ [panels (i) and (j)] for the Redfield regime (left column with $L=1000$) and the Landau-Zener regime (right column with $L=100$). Black closed circles denote the results based on the exact dynamics everywhere but on panel (b) and (d) where they are denoted by dashed lines. In the left column, the violet full lines denote non-Markovian RQME results and the violet dashed-dotted lines the definitions of heat and work via Eq.~(\ref{eq:30}). In the right column, the continuous time and discrete Landau-Zener predictions match exactly so that only the former are shown in green color. The vertical lines on panels (b) and (d) are plotted at multiples of the Heisenberg time $\tau_h = 2 \pi d$. Also $ \varepsilon_0 =5$, $\dot{\varepsilon}=1$, $\epsilon_c = 100$, $\nu = 0.1$, $\beta = 0.1$, and $\mu = 50$.}
\end{figure}

Figure~\ref{fig5} compares thermodynamic quantities calculated using the approaches described in the three previous subsections: the total system, the Redfield, and the Landau-Zener approach. The important observation is that the three approaches coincide very well for the integrated quantities: heat, work and entropy production.  

We now briefly describe the physics explaining the behavior of the heat and work. 
The total work is made of a mechanical work $W_{mech}$ and a chemical work $\mu N$ contribution. Since $\dot{\varepsilon} > 0$, the mechanical work is always positive. Initially, the dot energy is below the Fermi level $\mu$ and hence the fully occupied dot ($\ini = 1$) cannot easily transfer its electron to the reservoir. Hence until the energy of the dot $\varepsilon_t$ reaches the $k_B T$ vicinity of the Fermi level $\mu$, i.e. up to $t=40$, the mechanical work dominates and leads to an increase in total work $W$. In the $k_B T$ vicinity of $\mu$, the dot will transfer its electron to the reservoir, thus producing a negative chemical work which will decrease the total work $W$ ($|\mu N| > |W_{mech}|$) as one can see beyond $t=45$. The heat $Q$ is made of an energy current and a chemical work contribution. When the dot is deep in the Fermi sea, $\varepsilon_t < \mu$, the heat $Q$ grows and as it passes beyond the Fermi level the heat starts decreasing.

In the Redfield regime, we observe that the alternate definitions in Eq.~(\ref{eq:30}) are quite accurate for long times, but discrepancies are observed in the short-time limit as seen in the insets of Figs.\ref{fig5}(a) and (c). This is consistent with the fact that we expected these discrepancies in presence of non-Markovian effects. These discrepancies survive at longer times for the time-integrated quantities as seen in Figs.~\ref{fig5}(e), (g), and (i). The effect of the interaction energy $V$ in Eq.~(\ref{eq:24}) is thus not to be neglected. It reflects most significantly on the entropy production in Fig.~\ref{fig5}(i).

\begin{figure}[tb!]
\includegraphics[width=0.8\columnwidth]{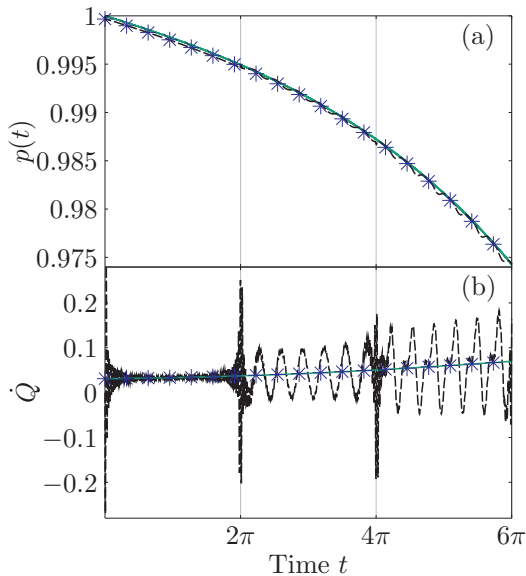}
\caption{\label{fig6} (Color Online) Occupation probability (panel a) and heat current (panel b) for the exact quantum dynamics (black dashed lines) [Eq.~(\ref{eq:3})], the discrete time LZQME (blue asterisk) [Eq.~(\ref{eq:11})], and the continuous time LZQME (green solid line) [Eq.~(\ref{eq:15})]. The parameters are the same as the Landau-Zener regime (right column) of Fig.~\ref{fig5} with an Heisenberg time $\tau_h=2\pi$. The discrete time LZQME points $p_n$ are located right before the avoided crossing $t_n$.}
\end{figure}

In the Landau-Zener regime the heat and work rate defined in the total system display oscillations as can be seen on Figs.~\ref{fig5}(b) and (d)]. These significantly increase after every multiple of the Heisenberg time $\tau_h = 2\pi d$. They can be seen as recurrences arising from the finiteness of the reservoir. Interestingly, these effects are nearly invisible at the level of the occupation probabilities $p(t)$, but they are much more pronounced on the heat and work rate as can be clearly seen in Fig.~\ref{fig6}. This figure also shows that the discrete LZQME (blue asterisk) predictions which iterate from crossing to crossing, match very well with the exact results. This is because the oscillations, and thus the coherent effects, only arise in between the avoided crossings. 



\section{Periodic driving}\label{sec:5}

\begin{figure}[tb!]
\includegraphics[width=\columnwidth]{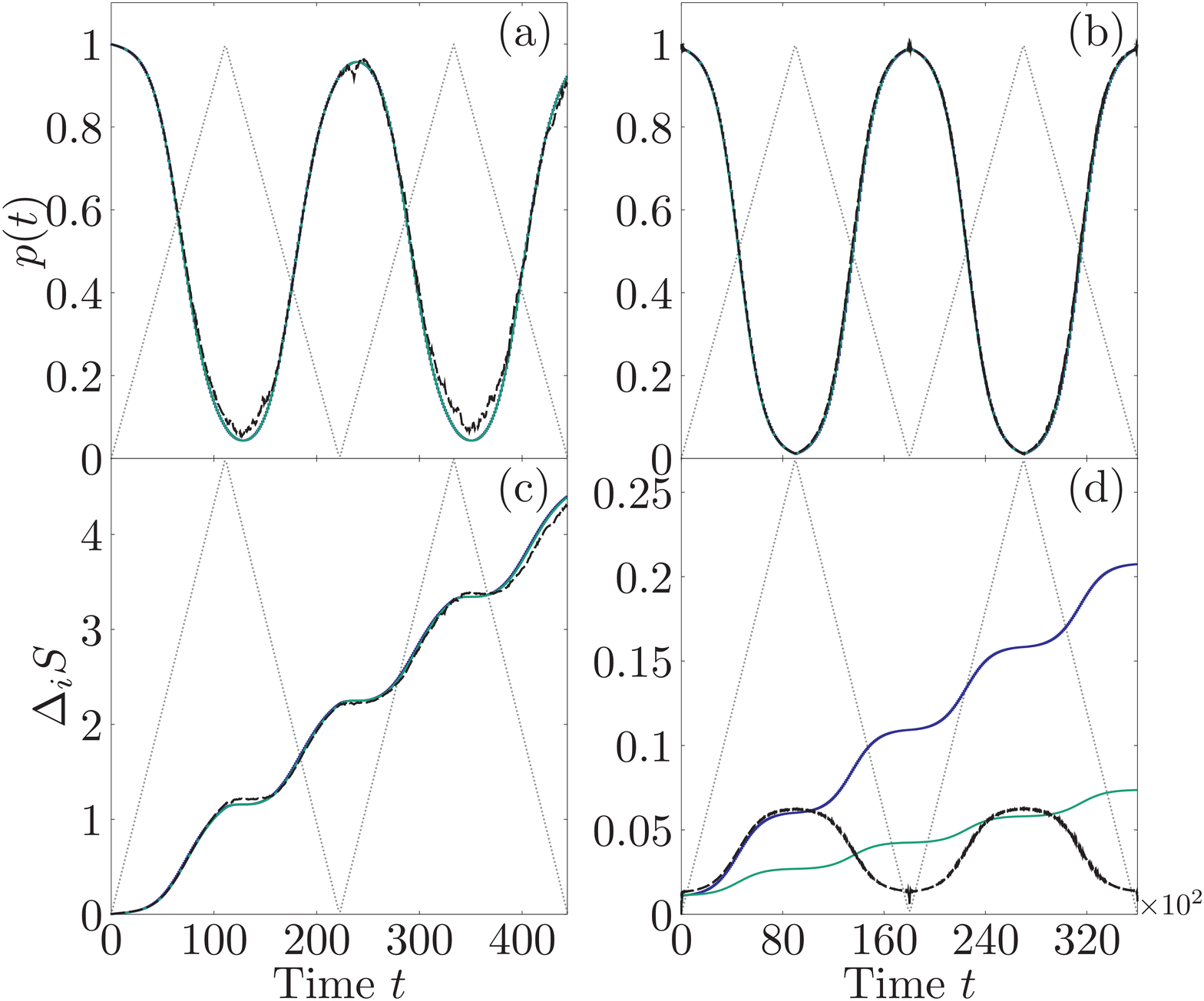}
\caption{\label{fig7} (Color Online) System occupation for the exact quantum dynamics (black dashed lines) [Eq.~(\ref{eq:3})], discrete time LZQME (blue jagged line) [Eq.~(\ref{eq:11})], and continuous time LZQME (green solid line) [Eq.~(\ref{eq:15})] for fast driving [panel (a), $\dot{\varepsilon}=0.81$] and slow driving [panel (b), $\dot{\varepsilon}=0.01$]. The corresponding entropy production is plotted in panels (c) and (d). The gray dotted line in the background illustrates the form of the driving between $\varepsilon_0$ and $\varepsilon_f$. Also $ \varepsilon_0 =5$, $L=100$, $\epsilon_c = 100$, $\nu = 0.1$, $\beta = 0.1$, and $\mu = 50$.}
\end{figure}

Until now we focused on linear drivings of the dot energy. In this final section we consider a periodic sequence of linear ramps as illustrated by the gray dotted line in Fig.~\ref{fig7}.

During the first upward drive (Fig.~\ref{fig7}, left column), the population and the entropy production of the exact quantum evolution agree well with the discrete as well as continuous LZQME. This is not surprising since the assumption that the driven dot interacts only with equilibrated reservoir levels is perfectly satisfied. However during the first downward drive, the dot may re-interact with reservoir levels with which it already interacted during the upward drive. This effect is not accounted for by the LZQME. At the level of the occupation this effect is negligible for fast driving as seen in Fig.~\ref{fig7}(a). Indeed, the dot only interacted with a small fraction of the reservoir levels along its way up, given by $(1-R)$. The probability to re-interact on the way down with precisely one of reservoir level that got affected on the way up is therefore very low. Obviously, as the number of cycles increase this probability will increase and discrepancies will start to occur. We also see in Fig.~\ref{fig7}(a) that the mismatch is more pronounced in the regions where the driving changes direction, probably due to interference effects caused by coherences which had no time to decay. In the case of very slow driving displayed in Fig.~\ref{fig7}(b), the dot interacts with every reservoir level and at every interaction they exchange their occupation. This means that the effect of the upward drive is to lift the entire occupation of the reservoir levels by one level, i.e. $f_{n+1} \to f_{n}$. The exchanges occurring during the downward drive will undo the effect of the upward exchanges and restore the dot and the reservoirs to their original occupation. Because during this downward drive the dot sees again essentially ``thermal'' reservoir levels (although they are slightly shifted by one level) as required by the LZQME, the agreement between the exact occupation and the LZQME prediction remains good as can be seen in Fig.~\ref{fig7}(b). At the level of the entropy production however, while the agreement remains good for fast driving it becomes completely wrong for slow driving as can be seen in Figs.~\ref{fig7}(c) and (d). Indeed, in this case the ramping down phase of the dynamics plays the role of a time reversal operation of the ramping up phase. It bring back the dot and the reservoir to their original states. The entropy production calculated with the exact dynamics nicely accounts for the reversibility of this process, as seen in Fig.~\ref{fig7}(d) [black dashed line]. The continuous time LZ theory (green solid curve) gives a completely wrong prediction since it inherently requires the diabatic approximation as stated above Eq.~\eqref{eq:12}. Whereas the entropy production from the discrete time LZ theory (blue jagged line), which is devoid of the diabatic approximation, is able to capture the entropy production on the upward drive correctly. Both LZ theories are unable to capture the reversibility in the entropy production as they assign a positive entropy production to every transition, assuming implicitly that the reservoir is always thermal and fresh of correlations.

\section{Conclusions and perspectives}\label{sec:6}

In this paper we investigated the validity of two kinetic schemes in predicting the correct dynamics and thermodynamics of a time-dependently driven quantum system interacting with a thermal reservoir. The first is the well known Redfield theory generalized to time-dependent drivings and valid for dense reservoirs. The second is a Landau-Zener scheme valid for sparse reservoirs.   

Our study relied on various simplifying assumptions: the total system is a noninteracting Fermionic system, the time-dependent driving is piece-wise linear and commutes with the system Hamiltonian at all times, and the case of multiple reservoirs has not been considered. Also, in the simulations the system-reservoir couplings were taken constant and the reservoir levels equally spaced. Which of these assumptions can be relaxed will be the object of future work.

Nevertheless, the detailed comparison between the kinetic predictions and the numerically exact dynamics revealed a number of interesting features. 
$-$ The driving tends to extend the range of validity of Redfield theory towards lower reservoir densities of state. 
$-$ Continuous time Landau-Zener theory and Redfield theory give rise to the same quantum master equation.  
$-$ The Landau-Zener regime is not incoherent per se, but rather system-reservoir coherences do not explicitly play a role at the level of the discrete time description between avoided crossings.
$-$ The kinetic-based thermodynamics can very well reproduce the exact thermodynamics identities derived in Ref.~\cite{Esposito2010}. For Redfield theory, beyond the Markov approximation it is important to include the system-reservoir interaction energy as part of the system energy. For Landau-Zener theory, the reservoir levels interacting with the system must always be thermal. 

The kinetic and thermodynamic description of driven open quantum systems still needs to be more systematically understood. In particular the key role that time-dependent driving seem to play on thermalization in open quantum systems \cite{Breuer2000, Ketzmerick2010, Shirai2015, Shirai2016, Iwahori2016}. 

\begin{acknowledgments}
Juzar Thingna and Massimiliano Esposito are supported by the European Research Council (Project No. 681456). Felipe Barra gratefully acknowledges the financial support of FONDECYT grant 1151390.
\end{acknowledgments}


\begin{thebibliography}{100}
\bibitem{note1}
The term mixing here implies that the bare system and reservoir energy levels due to the presence of a coupling are hybridized such that several reservoir levels form a hybrid level.
\end{thebibliography}
%
\end{document}